\documentclass[11pt,twoside]{article}

\usepackage{asp2006}
\usepackage{epsf}
\usepackage{lscape}

\begin{document}
\title{CARMENES: Calar Alto high-Resolution search for M dwarfs with Exo-earths
with a Near-infrared Echelle Spectrograph}   %%% Fill in title
\author{A.~Quirrenbach,\altaffilmark{1} 
P.~J.~Amado,\altaffilmark{2} 
H.~Mandel,\altaffilmark{1} 
J.~A.~Caballero,\altaffilmark{3} 
I.~Ribas,\altaffilmark{4}
A.~Reiners,\altaffilmark{5} 
R.~Mundt,\altaffilmark{6} 
M.~Abril,\altaffilmark{2} 
C.~Afonso,\altaffilmark{6} 
J.~L.~Bean,\altaffilmark{5} 
V.~J.~S.~B\'ejar,\altaffilmark{7} 
S.~Becerril,\altaffilmark{2} 
A.~B\"ohm,\altaffilmark{6} 
C.~C\'ardenas,\altaffilmark{2} 
A.~Claret,\altaffilmark{2} 
J.~Colom\'e,\altaffilmark{4}
L.~P.~Costillo,\altaffilmark{2} 
S.~Dreizler,\altaffilmark{5} 
M.~Fern\'andez,\altaffilmark{2} 
X.~Francisco,\altaffilmark{4}
R.~Garrido,\altaffilmark{2} 
J.~I.~Gonz\'alez~Hern\'andez,\altaffilmark{3} 
E.~W.~Guenther,\altaffilmark{8} 
J.~Guti\'errez-Soto,\altaffilmark{2} 
V.~Joergens,\altaffilmark{6} 
A.~P.~Hatzes,\altaffilmark{8} 
T.~Henning,\altaffilmark{6} 
E.~Herrero,\altaffilmark{4}
M.~K\"urster,\altaffilmark{6} 
W.~Laun,\altaffilmark{6} 
R.~Lenzen,\altaffilmark{6} 
U.~Mall,\altaffilmark{6} 
E.~L.~Mart\'{\i}n,\altaffilmark{7} 
S.~Mart\'{\i}n-Ruiz,\altaffilmark{2} 
D.~Montes,\altaffilmark{3} 
J.~C.~Morales,\altaffilmark{4}
R.~Morales~Mu\~noz,\altaffilmark{2}
A.~Moya,\altaffilmark{2}
V.~Naranjo,\altaffilmark{6} 
O.~Rabaza,\altaffilmark{2}
A.~Ram\'on,\altaffilmark{2}
R.~Rebolo,\altaffilmark{7} 
S.~Reffert,\altaffilmark{1} 
F.~Rodler,\altaffilmark{7} 
E.~Rodr\'{\i}guez,\altaffilmark{2}
A.~Rodr\'{\i}guez~Trinidad,\altaffilmark{2}
R-.R.~Rohloff,\altaffilmark{6} 
M.~A.~S\'anchez~Carrasco,\altaffilmark{2}
C.~Schmidt,\altaffilmark{5} 
W.~Seifert,\altaffilmark{1} 
J.~Setiawan,\altaffilmark{6} 
O.~Stahl,\altaffilmark{1} 
J.~C.~Su\'arez,\altaffilmark{2}
G.~Wiedemann,\altaffilmark{9}
C.~del~Burgo,\altaffilmark{10}
D.~Galad\'{\i},\altaffilmark{11}
E.~S\'anchez-Blanco,\altaffilmark{12}
W.~Xu\altaffilmark{13}}   %%% Fill in author names
\altaffiltext{1}{Landessternwarte K\"onigstuhl, Zentrum f\"ur Astronomie Heidelberg, K\"onigstuhl 12, 69117 Heidelberg, Germany}
\altaffiltext{2}{Instituto de Astrof\'{\i}sica de Andaluc\'{\i}a (CSIC), Apartado 3004, 18080 Granada, Spain}
\altaffiltext{3}{Departamento de Astrof\'{\i}sica y Ciencias de la Atm\'osfera, Universidad Complutense de Madrid, 28040 Madrid, Spain}
\altaffiltext{4}{Institut de Ci\`encies de l'Espai (CSIC-IEEC), Campus UAB, Facultat de Ci\`encies, Torre C5, parell, 2a, 089193 Bellaterra, Barcelona, Spain}
\altaffiltext{5}{Institut f\"ur Astrophysik, Georg-August-Universit\"at, Friedrich-Hund-Platz 1, 37077 G\"ottingen, Germany}
\altaffiltext{6}{Max-Planck-Institut f\"ur Astronomie, K\"onigstuhl 17, 69117 Heidelberg, Germany}
\altaffiltext{7}{Instituto de Astrof\'{\i}sica de Canarias, V\'{\i}a L\'actea s/n, 38205 La Laguna, Tenerife, Spain}
\altaffiltext{8}{Th\"uringer Landessternwarte Tautenburg, Sternwarte 5, 07778 Tautenburg, Germany}
\altaffiltext{9}{Hamburger Sternwarte, Gojenbergsweg 112, 21019 Hamburg, Germany}      
\altaffiltext{10}{UNINOVA - Instituto de Desenvolvimento de Novas Tecnologias, Lisboa, Portugal}      
\altaffiltext{11}{Centro Astron\'omico Hispano-Alem\'an de Calar Alto, Almer\'{\i}a, Spain}      
\altaffiltext{12}{Dise\~no de Sistemas \'Opticos, Sevilla, Spain}      
\altaffiltext{13}{Wenli Xu Optical System Engineering, Spechbach, Germany}      
%%% Fill in author affiliations

\begin{abstract} %%% Abstract to run on from here.
CARMENES, {\em Calar Alto high-Resolution search for M dwarfs with Exo-earths
with a Near-infrared Echelle Spectrograph}, is a study for a next-generation instrument for
the 3.5\,m Calar Alto Telescope to be designed, built, integrated, and operated
by a consortium of nine German and Spanish institutions.
Our main objective is finding habitable exoplanets around M dwarfs, which will
be achieved by radial velocity measurements on the m\,s$^{-1}$ level in the
near-infrared, where low-mass stars emit the bulk of their radiation.
\end{abstract}

%%% MAIN BODY OF TEXT GOES HERE.
\section{Introduction}

So far, radial velocity exoplanet searches have mainly focused on
Solar-like main sequence stars. However, searches for exoplanets around M dwarfs have also been successful, and knowing the frequency of these objects places important constraints on planet formation scenarios.
Some of the least massive exoplanets known orbit low-mass M dwarfs
(e.g.\ {GJ\,581} and {GJ\,876} -- Rivera et~al. 2005; Udry et~al. 2007; Mayor
et~al. 2009).

M dwarfs are the most common stars in the solar neighborhood, and their
habitable zones lie close to them. This means that ``habitable'' planets around M dwarfs have large radial-velocity signatures and a large probability of showing transits.
In spite of their interest, M dwarfs have not been searched for planets as
extensively as late-F, G, and K stars, because of their faintness in the
optical, where most radial velocity searches are being performed. 
A near-infrared spectrograph with a radial velocity accuracy on the m\,s$^{-1}$
level would be more efficient to detect Earth-like planets around stars with
spectral types later than about M3. Radial velocities measured in the near infrared would also be less susceptible to stellar radial-velocity noise. As a high-resolution near-infrared spectrograph dedicated to a planet survey does not exist yet, we have performed a study of such an instrument for the Calar Alto Astronomical Observatory.

\section{CARMENES Design Overview}

The CARMENES study was initiated as a joint Spanish-German answer to a call for
ideas for new instruments for Calar Alto.
In this contribution, we describe a summary of the CARMENES configuration as
it was presented in the Conceptual Design Review in early October 2009. (The
configuration may have changed since then; visit our webpage\footnote{\tt
http://www.ucm.es/info/carmenes.} for later developments). CARMENES is expected to become operational in 2013.

CARMENES will be fiber-fed from a front-end at the prime focus of the 3.5\,m
Calar Alto Telescope.
The three cross-dispersed echelle spectrograph channels (NIR, VIS, MOS; see
below) will be located in the coud\'e room of the telescope, where they can be thermally and mechanically stabilized.
The instrument will cover from 500 to 1800\,nm in one shot with a near-IR radial velocity
precision requirement of 3\,m\,s$^{-1}$.

Thanks to an optical design with a mosaic of two 2\,k\,$\times$\,2\,k detectors
and an R2.9 echelle grating, the near-infrared (NIR) channel will cover from 950\,nm
($Y$ band) to about 1800\,nm ($H$ band) with a spectral resolution R = 85\,000
in 31 echelle orders.
The cross disperser will consist of two S-NPH2 prisms.
Most of the optomechanical components of the NIR channel will be located inside
a vacuum tank at $T$ = --30$^\circ$C. The NIR channel needs an image slicer and an image scrambler.

The design of the visible (VIS) channel, which will cover from 500 to 900\,nm
with R = 60\,000 in 42 echelle orders, will be based on the successful FEROS 
instrument. 
Simultaneous observations with the NIR and VIS channels will allow us to monitor
the main activity indicators (H$\alpha$ and the calcium triplet) with the same
temporal sampling as the radial-velocity curve, which will help us 
discriminate between activity-induced and planet-induced radial-velocity
variations.

The multiobject (MOS) visible channel takes advantage of the $\sim$0.8\,deg$^2$ field of the 3.5\,m telescope
to acquire the spectra of $\sim$12 bright stars during the
M dwarf observations; in this way a survey of G and K giants can be conducted without any additional telescope time.

CARMENES may share the telescope prime focus with another instrument. 
The design therefore foresees a common front end mounted at the primary focus
behind the K3 corrector. 
This front end will contain the fiber positioners for the CARMENES
NIR, VIS and MOS channels. Our primary channels (NIR and VIS) require two fibers
each: one for the object and one for a ThAr lamp or sky.

\acknowledgements
%%% Text of acknowledgements runs on after this command.
The CARMENES study was funded by the Centro Astron\'omico Hispano-Alem\'an,
which is operated jointly by the Max-Planck-Institut f\"ur Astronomie
(Max-Planck-Gesellschaft) and the Instituto de Astrof\'{\i}sica de
Andaluc\'{\i}a (Consejo Superior de Investigaciones Cient\'{\i}ficas).

%%% THE BIBLIOGRAPHY

\end{document}